\documentstyle[12pt,aasms4]{article}
\lefthead{Chakrabarti and Manickam}
\righthead{QPO of GRS1915+105}
\def\lsim{\lower.5ex\hbox{$\; \buildrel < \over \sim \;$}}
\def\gsim{\lower.5ex\hbox{$\; \buildrel > \over \sim \;$}}

\begin{document}

\title{Correlation among QPO frequencies and
Quiescence-state Duration in Black Hole Candidate\\ 
GRS 1915+105}

\author{Sandip K. Chakrabarti and Sivakumar G. Manickam} 
\affil{S. N. Bose National Centre for Basic Sciences, JD-Block, Salt Lake,
Calcutta, 700091, INDIA\\
e-mail: chakraba@boson.bose.res.in \& sivman@boson.bose.res.in}

\begin{abstract}
We discover a definite correlation between the frequency of the
quasi-periodic oscillations (QPO) in quiescence states and the duration of the quiescence 
state of the transient X-ray source GRS 1915+105. We find that
while the QPO frequency can be explained with the 
oscillation of shocks in accretion flows, the switching of burst
to quiescence states (and vice versa) and their duration can be explained by
assuming an outflow from the post-shock region. The duration
of the quiescence state is inversely related to the
QPO-frequency. We derive this relation. We also find the correlation between 
the observed low ($\sim 0.001-0.01$Hz)
and the intermediate ($1-10$Hz) QPO frequencies.  Our analytical solutions are verified
by analyzing several days of public-domain data from RXTE.
\end{abstract}


\section{Introduction}

X-ray transient source GRS 1915+105 in our galaxy exhibits various types 
of quasi-periodic oscillations with frequencies ranging from $\sim 0.001-0.01Hz$
to $\sim 67Hz$ (Morgan et al, 1997; Paul et al. 1998;
Yadav et al. 1999). The object is
sometimes in a flaring state with regular and quasi-regular bursts 
and quiescences, while at some other time it is in usual low-hard and
high-soft states. While the light curves look very 
chaotic with no apparent similarity between observations in two different
days, some of the features are classifiable: (a) low-frequency 
QPO ($\nu_{L}\sim 0.001-0.01$Hz) is due to the transition between
burst and quiescence states (which we term as `on'-state and `off'-state respectively) 
and vice versa; (b) the intermediate frequency
QPO ($\nu_{I}\sim 1-10$Hz) could be due to oscillations 
of shocks located at tens to hundreds of Schwarzschild radii $R_g$ ($=2GM/c^2$
is the Schwarzschild radius. Here, $M$ is the mass of the black hole, $G$, and $c$ are the 
gravitational constant and velocity of light respectively) and 
(c) very high frequency QPO ($\nu_{H}\sim 67$Hz), if at all present,
could be due to oscillations of the shocks located at several $R_g$. 
$\nu_I$ is generally observed during quiescence states. Typically,
a shock located at $R_s$ (unless mentioned otherwise, measured hereafter in units of $R_g$), 
produces an oscillation of frequency,
$$
\nu_{I} = \frac{1}{t_{ff}} \sim \frac{1}{R} R_s^{-\alpha} \frac{c v_0}{R_g} \ \ s
\eqno{(1)}
$$
where, $R$ is the compression ratio of the gas at the shock. Here we used 
a result of Molteni, Sponholz \& Chakrabarti (1996, hereafter referred to as 
MSC96) which states that the time-period of QPO oscillation is comparable to
the infall time ($t_{infall} = t_{ff} R \sim R_s^{3/2}$)
in the post-shock region. However, we assume now that the post-shock 
velocity is not necessarily $R_s^{-1/2}$ dependent as in a free fall but could 
be slowly varying, especially when angular momentum is high.
In this case, $t_{infall} \propto R_s^\alpha$. Clearly, $\alpha\sim 3/2$ 
for a low angular momentum freely falling matter and $\alpha\sim 1$ for 
a post-shock flow of constant velocity $v_0 c/R$. 
Here $v_0$ is a dimensionless quantity which is exactly unity for a 
free-fall gas. For a gas of $\gamma=4/3$, $R \sim 7$ 
and for $\gamma=5/3$, $R\sim 4$, when the shock is strong. Thus, 
for instance, for a $\nu_I=6$Hz, $R_s \sim 38$ for $M=10M_\odot$ and 
$\gamma=4/3$. For $\nu_H=67$Hz, $R_s \sim 8$ for the same parameters. 
MSC96 and Chakrabarti \& Titarchuk (1995, hereafter CT95) postulated that since black hole QPOs
show a large amount of photon flux variation, they cannot be explained
simply by assuming some inhomogeinities, or perturbations in the flow.

If the QPOs are really due to shock oscillations, they should almost 
disappear at low energy soft X-rays, since these X-rays are produced
in pre-shock flow which does not participate in large-scale oscillations.
Second, a shock-compressed gas with compression ratio $R>1$, must
produce outflows or extended corona which pass through sonic points located at
at $R_c = f_0 R_s/2$, where $f_0=R^2/(R-1)$, if the flow is assumed 
to be isothermal till $R_c$ (Chakrabarti 1998, 1999, hereafter C98 and C99
respectively). In this solution the location of the sonic point $R_c$ and
ratio between outflow and inflow rates are functions of the 
compression ratio $R$ of the shock alone. Till the sonic point $R_c$,
matter is subsonic and this subsonic volume is filled in a time of (C99),
$$
t_{fill}=\frac{4\pi R_c^3 <\rho>}{3 {\dot M}_{out}} ,
\eqno{(2)}
$$
where, $<\rho>$ is the average density of the sonic sphere, and ${\dot M}_{out}$ is the
outflow rate. The Compton cooling becomes catastrophic 
when $<\rho>R_c k_{es} \gsim 1$,  $k_{es}=0.4$ is the Thomson scattering 
opacity. Thus the duration of the off-state (i.e., duration between 
the end of a burst and the beginning of the next burst) is given by,
$$
t_{off} = \frac{4\pi R_c^2}{3 {\dot M}_{out} k_{es}}.
\eqno{(3)}
$$
We use now a simple relation between inflow and outflow rates
given by (C98, C99),
$$
\frac{{\dot M}_{out}}{{\dot M}_{in}} = R_{\dot m}=
\frac{\Theta_{out}}{\Theta_{in}}\frac{R}{4} 
[\frac{R^2}{R-1}]^{3/2} exp  (\frac{3}{2} - \frac{R^2}{R-1})
\eqno{(4)}
$$
where, $\Theta_{in}$ and $\Theta_{out}$ are the solid angles of the inflow 
and the outflow respectively. Because of the uncertainties 
in $\Theta_{in}$, $\Theta_{out}$ and ${\dot M}_{in}$ (subscript `in' 
refers to the accretion rate) we define a dimensionless parameter,
$$
\Theta_{\dot M} = \frac{\Theta_{out}}{\Theta_{in}} \frac{{\dot M}_{in} } {{\dot M}_{Edd}}.
\eqno{(5)}
$$
where, ${\dot M}_{Edd}$ is the Eddington rate. 
Using Eqs. (4-5), we get the following expression for $t_{off}$ as,
$$
t_{off}=\frac{10.47} {(R-1)^{1/2}} \frac{R_s^2 R_g^2 exp(f_0-\frac{3}{2})} {{\dot M}_{Edd} \Theta_{\dot M}} \ \ s .
\eqno{(6)}
$$
Or, eliminating shock location $R_s$ using eq. (1) and $\alpha=3/2$, $v_0=1$, we obtain,
$$
t_{off}= 14.1 \frac{exp(f_0-\frac{3}{2})} {R^{4/3} (R-1)^{1/2}\Theta_{\dot M}} 
(\frac{M}{10 M_\odot})^{-1/3} \nu_I^{-4/3} s
\eqno{(7)}
$$
For an average shock of strength  $2.5\lsim R \lsim 3.3$, 
the result is insensitive to the compression ratio. Using 
average value of $R=2.9$ and for $\Theta_{\dot M} \sim 0.1$ 
(which corresponds to $0.1$ Eddington rate for $\Theta_{out} 
\sim \Theta_{in}$) we get,
$$
t_{off}=461.5 
(\frac{0.1}{\Theta_{\dot M}}) (\frac{M}{10M_\odot})^{-1/3} \nu_I^{-4/3} s.
\eqno{(8)}
$$
Thus, the duration of the off-state must go down rapidly
as the QPO frequency increases if the flow geometry and the net accretion
rate remains fixed. When one considers a constant velocity post-shock flow,
$\alpha=1$, and $v_0=0.066$ 
(chosen so as to keep the same numerical coefficient 
as in eq. 8) the above equation is changed to,
$$
t_{off}=461.5 (\frac{0.1}{\Theta_{\dot M}}) (\frac{M}{10M_\odot})^{-1} (\frac{v_0}{0.066})^2\nu_I^{-2} s.
\eqno{(9)}
$$
Interestingly, $v_0=0.066$ (i.e., a constant velocity of seven percent of 
the velocity of light) is very reasonable for a black hole accretion.

If the hot post-shock gas of height $\sim R_s$ 
intercepts $n$ soft photons per second,
from the pre-shock Keplerian component (CT95),  
it should intercept about $nf_0^2/4$ soft photons per second
when the sonic sphere of size $R_c$ is filled in. Thus, the
photon flux in the burst state should be about $f_0^2/4 \gsim  4$ 
times larger compared to the photon flux in off-state. Depending
on the degree of flaring, and the fact that the wind is bent backward 
due to centrifugal force, the interception may be higher.

Since $\nu_L$ is basically due to recurrences of on and off-states,
it is clear that $\nu_L \sim 1/<t_{off}+t_{on}>$. Here, $t_{on}$ is the 
duration of the burst state which may be very small for extremely regular 
(spiky) bursts  reported by in Taam, Chen \& Swank (1997) and Yadav et al. 1999. In this case, 
$$
\nu_L=0.0022 (\frac {\Theta_{\dot M}}{0.1}) (\frac{10M_\odot}{M})\nu_I^{2} \ \ \rm{Hz}.
\eqno{(10)}
$$
When on-state has a non-negligible duration ($t_{on}\ne 0$), it is found to be directly related to the 
$t_{off}$ (Belloni et al., 1997; Yadav et al. 1999). Assuming $t_{on}\sim t_{off}$, the $\nu_L$ 
would be less by a factor of two when on states are broad. When the burst 
is very regular but `spiky' (i.e., with momentary on-state), $t_{on}\approx0$. 
The presence of $\nu_L$ for these regular `spiky' bursts are reported in Manickam
and Chakrabarti (1999a)

Thus, if our shock oscillation solution for QPO is correct, the observations
must pass all the following tests: (a) the QPO in the off-state
must disappear at low energies, (b) the QPO must 
generally be absent in the on-state, when the sonic sphere is
cooled down, (c) the intermediate QPO frequency must be correlated with $t_{off}$ as in Eqs. (8-9)
and (d) the photon flux must jump at least a factor of $4$ or more
when going from quiescence to burst state. In addition, (e) lowest frequency $\nu_L$ observed
must be correlated to the intermediate  QPO frequency $\nu_I$ by  eq. (10). There are 
uncertainties regarding the inflow velocity and actual volume-filling time, but we expect that 
above relations to be satisfied in general.

In the present {\it Letter} we show that observations do pass through these
tests and therefore the shock oscillation model may be the correct picture.
In the next Section, we present detailed analysis of 
some of the observational results on GRS1915+105  available in public archive
and show how they point to the shock oscillation model. Finally in
\S 3, we make concluding remarks.

\section {Observational Results}

Figure 1 shows a light curve of the first phase of observation
of June 18th, 1997, on the right panel. The average count rate (per second)
vary from around $\sim 5000$ in the off-state
to about $\sim 24,000$ in the on-state. The ratio
of the fluxes is about $\sim 5$. The duration 
of these states vary chaotically. At the mean location
of a few off-states (arrows on right axis), observation time is marked in seconds. For each
of these off-states, the power density spectrum (PDS) in arbitrary
units is drawn in the left panel. The most prominent QPO frequencies
($\nu_I$ in our notation) are connected by a dashed 
curve just to indicate its variation with time. There are 
some weaker peaks which follow the short dashed curve, indicating that they may be 
higher harmonics. Observations of this kind for several other days show similar
variations in QPO frequencies and details are presented
elsewhere (Manickam \& Chakrabarti, 1999ab).

In Figure 2, variation of $\nu_I$ with the duration $t_{off}$ of the off-states
(triangles) for the whole observation period on June 18th, 1997 in the log-log scale. 
Observational results from several other days (May 26th, 1997; June 9th, 1997; June 25,
1997; October 7th, 1996; October 25th, 1996) are also plotted on the same curve with circles, filled
squares, squares, filled circles and stars respectively. We did not put error bars
since in duration scale is it uniformly $\pm 2$ seconds, and in frequency scale
error bar is decided by the chosen bin-size while obtaining the PDS
(In our analysis it remains around 0.15-0.25 Hz, increasing monotonically with
QPO frequency). Times at which the photon flux is halved during the
on-to-off and off-to-on transitions are taken respectively to be the beginning
and the end of an off-state. Equation (9) is plotted in dashed 
lines with (from uppermost to the lowermost line) $\Theta_{\dot M} 
= 0.0034$, $0.0123$, $0.0163$, $0.028$, $0.0293$ and $0.043$
respectively indicating a slow variation of the accretion rate, provided the collimation property
remains the same. Two dotted curves, on the other hand, represent eq. (8), with
$\Theta_{\dot M}=0.06$ (top) and $0.093$ (bottom) respectively. We find 
that the inverse-squared law (eq. 9) may be a better fit 
to the observations. Since on Oct. 7th, 1996
the points are lumped closed to the lower right corner, not much could be said
about whether it follows our relation or not, but it is to be noted that 
its general behaviour (low frequency, high duration) follows our result 
for any reasonable $\Theta_{\dot M}$.  

Table 1 shows the variation of the $\nu_I$ (taken from Fig. 2) with days of observations. 
In 3rd column, the expected $\nu_L$ has been put ($\nu_L/2$ for Oct. 7 and June 18 
results as they show $t_{on}\approx t_{off}$). In 4th Column, the observed $\nu_L$ is
given. Generally, what we observe is that,
when the drift of $\nu_I$ is large, PDS around $\nu_L$ is also broad 
In any case, observed $\nu_L$ agrees with our expectations. 

\begin{figure}
\vbox{
\centerline{ TABLE 1 }
\centerline{\small\small  Correlation Between Low And Intermediate Frequencies of QPO (in Hz)}}
\begin{center}
\begin{tabular}{lllll}
\hline
\hline
Day & $\nu_I$ & $\nu_L$ from Eq. 10 & $\nu_L$ observed & Remarks \\
\hline
Oct. 7, 1996 &1.7-1.90 &0.0013-0.0016&0.001&broad on\\
Oct. 25, 1996 &2.9-3.5 &0.0006-0.0009&-----& sparse data\\
May 26, 1997 & 5.32-6.54 &0.0076-0.0115& 0.01 & spiky\\
June 9 1997 &6.22-7.42&0.0138-0.0197& 0.014& spiky\\
June 18 1997 &4.0-7.51&0.003-0.011& 0.01&broad on\\
June 25 1997 &6.61-7.42 &0.016-0.02& 0.022&spiky\\
\hline
\end{tabular}
\end{center}
\end{figure}

Figure 3 shows the PDS of the off-state centered at 1576s
(see Fig. 1) of the June 18th observation. The energy range  is given in each panel
Clearly, QPO disappears completely at low energies, exactly as is expected in the
shock oscillation model (MSC96) though QPO frequencies, when present, seem to be
energy independent (see, Manickam \& Chakrabarti, 1999b for details). The pre-shock flow
which emits soft radiation participates little in the oscillation as the
{\it fractional change} in the Keplerian disk due to shock oscillation  is negligible.
On the contrary, the {\it fractional change} in the size of the post-shock 
flow during the oscillation is large. Thus, the flux of hard X-rays oscillates as the size
of the post-shock region oscillates. This is the cause of QPO in our model.
In on-states, when they exist, the QPO is found to be very weak. 

\section{Discussion and Conclusions}

In this {\it Letter}, we have discovered a relation between the QPO frequency
in 1-10Hz range with the duration of the quiescence state at which the QPO is observed.
We also derived a relation between the low QPO frequency and the intermediate QPO frequency.
We analyzed several days of RXTE observations and showed that our relations are 
satisfied, especially when the average bulk velocity in the post-shock region is constant. 
We showed that the QPO disappears in the low energy, but is very strong in high energies. 
The photon flux is found to fluctuate, typically by a factor of $4$ or more, indicating that 
a vertically inflated post-shock region is responsible for interception of the soft photons
from a Keplerian disk. This factor seems to be similar to $f_0^2/4$ (C98, C99) for
any reasonable compression of the gas, which strengthens our belief that the
quasi-periodic cooling of the sonic sphere of the outflow from the post-shock
region may be responsible for the rapid transitions between on and off-states.
Our computation of the duration of the off-states from this considerations are found to be
quite reasonable. We find that for $\nu^{-4/3}$ law, $t_{off}$ is 
insensitive to the mass of the black 
hole while for $\nu^{-2}$ law, $t_{off}$ is inversely proportional to the mass.
Trudolyubov et al. (1999) recently found the duration of `hard' states varies as $-7/3$ power 
of the {\it lowest centroid frequency} for a {\it group} of data while we 
find an inverse-squared law when we choose the QPO frequency where the power is strongest. 

Although we chose a specific model for the outflow (C99) for concreteness,
the physical processes invoked are generic and the explanation should be
valid even when other models for outflows are used (except self-similar models). The 
shock location near the inner edge of the Keplerian disk can drift on viscous time scale
(see appendix of CT95 where the transition from Keplerian to sub-Keplerian
is plotted as a function of viscosity). The shocks can evacuate the disk, and form once again,
very similar to what was seen in the numerical simulation of Ryu et al. (1997). 
This drift would cause a drift in frequency as Trudolyubov et al. (1999) recently showed
(see also, Belloni et al. 1997; Markwardt, Swank \& Taam 1999, Muno et al. 1999). 
Our model invokes also outflows (which we believe form naturally in the
post-shock region and from the transition region from Keplerian to a sub-Keplerian flow)
which we find useful to explain variations in photon 
counts between `off' and `on' states.

\acknowledgments

This work is partly supported by a project (Quasi Periodic Oscillations in Black Hole Candidates)
funded by Indian Space Research Organization (ISRO).
The authors thank NASA for making RXTE data available and ISRO for creating a Data Bank
at their Centre where these data are stored.

{}

\vfil\eject
\begin{figure}
\caption {Plot of the light curve (right panel) and evolution of power density spectrum (left panel) of 
the first phase of June 18th, 1997 observation.
Off-states analyzed are marked by the time of observations on the right
axis. The QPO frequencies (where the power is strongest) 
are connected by a dashed curve to highlight the evolution of $\nu_I$ with time. }
\end{figure}

\begin{figure}
\caption {Variation of QPO frequency $\nu_I$ with duration of quiescence
states $t_{off}$. Dotted curves are the $t_{off} \propto \nu^{-4/3}$ law (eq. 9) derived using simple
free-fall velocity assumption. Dashed curves are the $t_{off}\propto \nu^{-2}$ (eq. 10)
with constant post-shock velocity law. The general agreement strongly points to the shock oscillation model.}
\end{figure}

\begin{figure}
\caption{Power density spectrum of the off-state (duration $87$s) 
centered at $1576$s constructed from selected channel intervals of the
binned PCA data. QPO is seen only in high energies, 
strongly pointing to the shock oscillation model.
}
\end{figure}
 
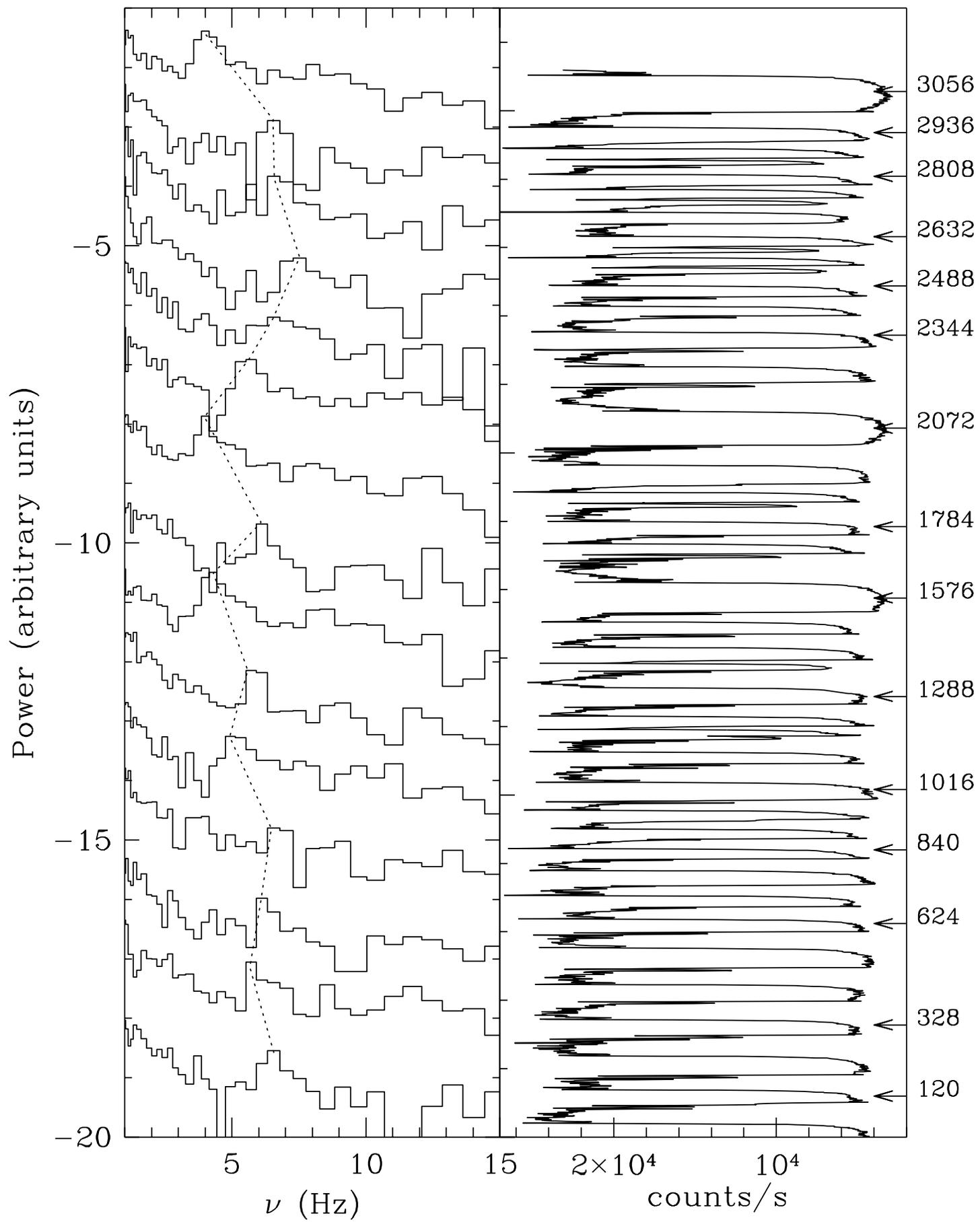
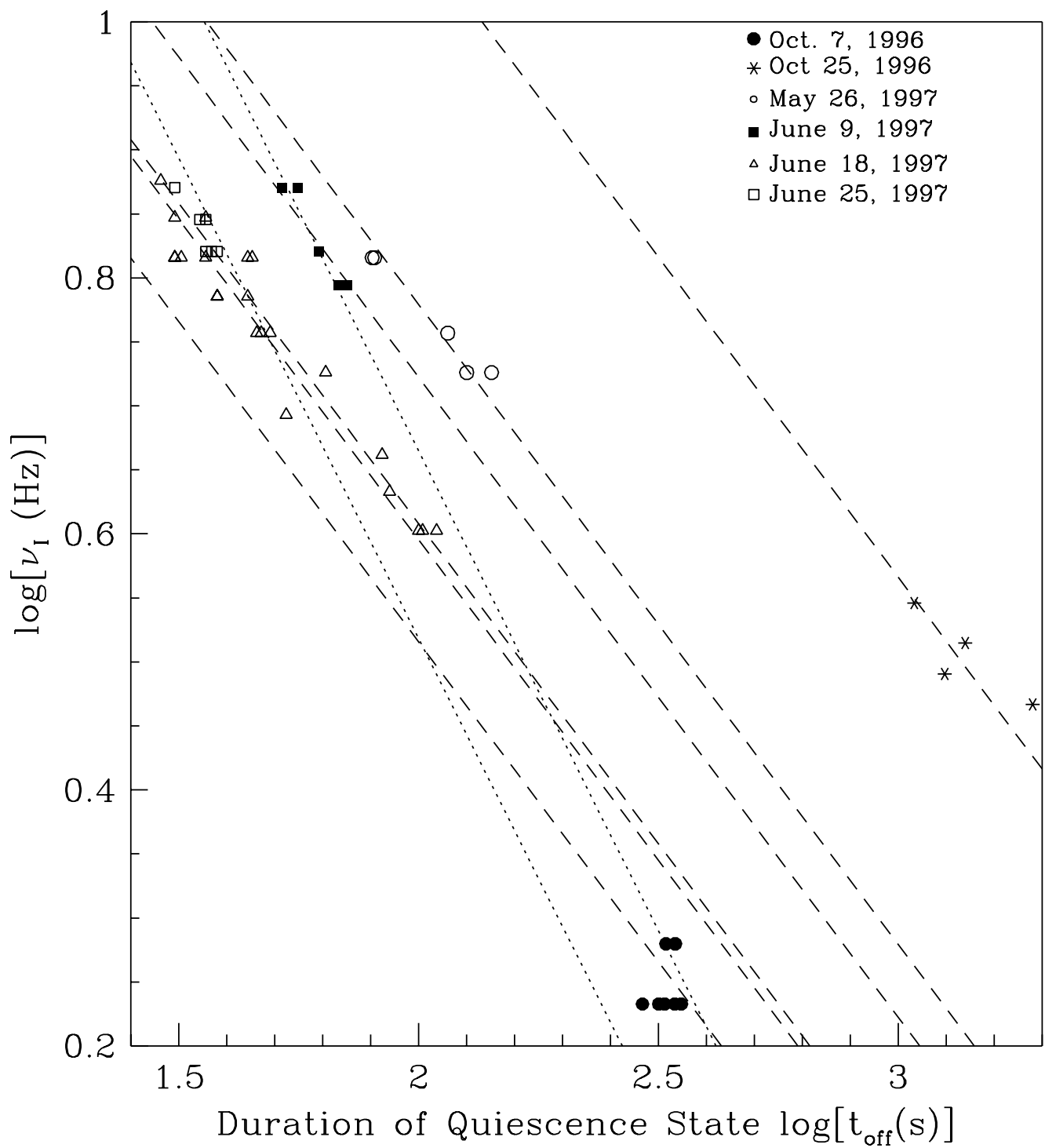
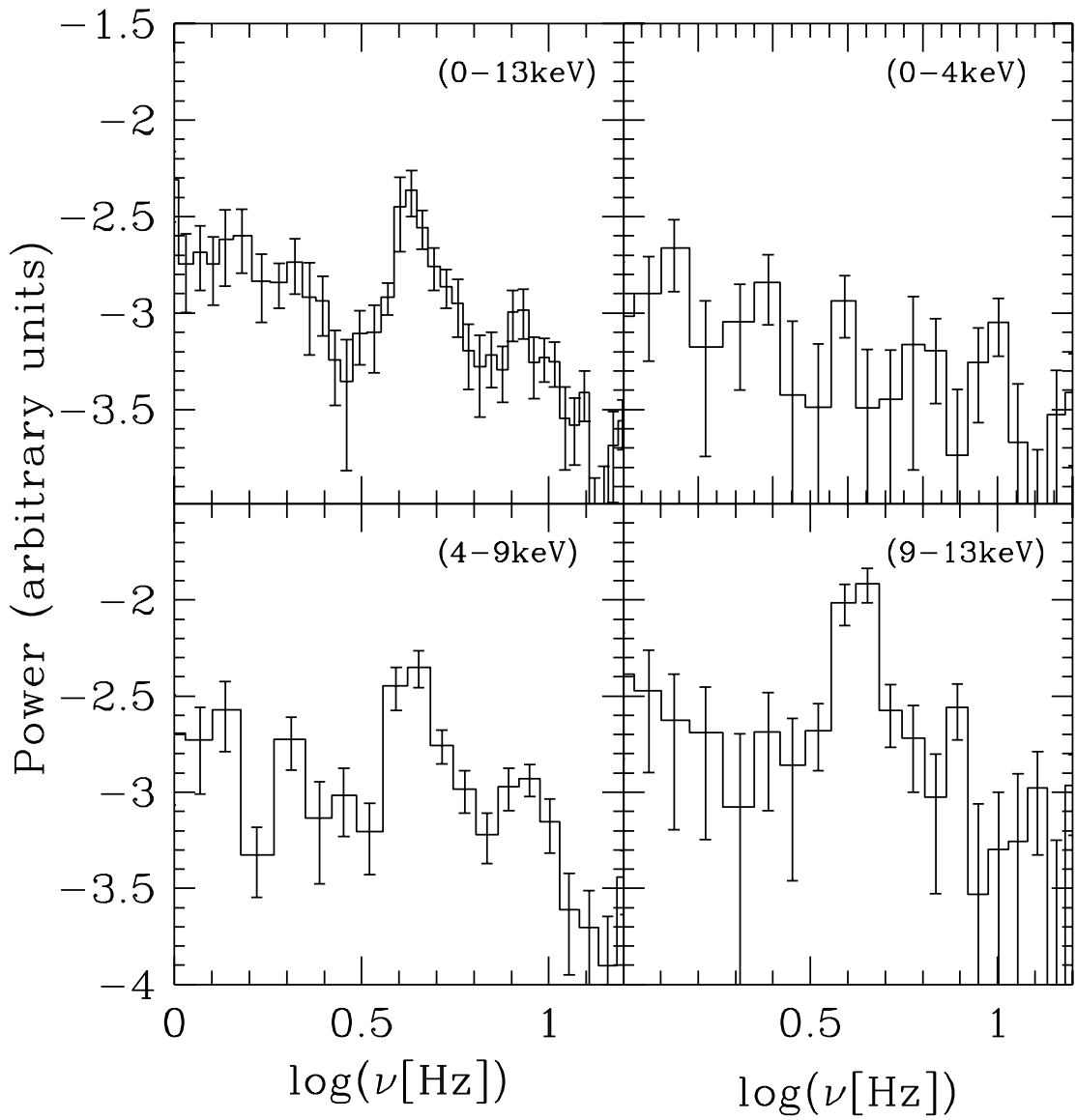
\end{document}